\documentstyle[aps,epsfig,twocolumn]{revtex}

\begin{document}

\newcommand{\ba}{\begin{eqnarray}}
\newcommand{\ea}{\end{eqnarray}}

\twocolumn[ \hsize\textwidth\columnwidth\hsize\csname@twocolumnfalse\endcsname 

\draft 
\title{Effect of the pseudogap on the mean-field magnetic penetration depth 
of YBa$_{2}$Cu$_{3}$O$_{7-\delta}$ thin films} 
\author{Brent R. Boyce, Kathleen M. Paget, Thomas R. Lemberger}
\address{Dept. of Physics, Ohio State University, Columbus, Ohio 43210-1106}
\date{\today}
\maketitle 

\begin{abstract}

We report measurements of the $ab$-plane penetration depth, $\lambda (T)$, in 
YBa$_{2}$Cu$_{3}$O$_{7-\delta}$ films at various $\delta$. At optimal doping, critical fluctuation 
effects are absent, and $1/\lambda^{2}(T)$ from 4 K to 0.99 $T_{C}$ is that of a clean, 
strong-coupling d-wave superconductor with $\Delta_{0}(0)/k_{B}T_{C} \simeq 3.3$.  As in crystals, 
underdoping reduces the superfluid density, $n_{s}(0) \propto 1/\lambda^{2}(0)$, without affecting 
the low-$T$ slope of $1/\lambda^{2} (T)$.  These results, as well as electronic heat capacity data, 
are well described by an {\it ad hoc} model in which contributions to the superfluid and entropy 
are lost from regions of the Fermi surface occupied by the pseudogap. 

\end{abstract}
 ]


A large body of experimental evidence indicates the opening of a {\bf k}-dependent gap, or 
pseudogap, at a temperature, $T^{*}$, above the superconducting transition temperature, $T_{C}$, in 
underdoped cuprates \cite{timusk}.  The pseudogap competes with the superconducting gap:  $T^{*}$ 
and the fraction of the Fermi surface (FS) occupied by the pseudogap increase with underdoping, 
while $T_{C}$, the superfluid density in the $ab$-plane, $n_{S}(0)$, and the peak value of the 
electronic specific heat coefficient, $\gamma(T)$, at $T_{C}$ decrease.  A great deal of effort 
currently focuses on understanding the coexistence of these two gaps.  Lee and coworkers propose 
that the fundamental physics lies in spin-charge separation, a key element being the segmentation 
of the FS into regions either occupied or unoccupied by the pseudogap in the normal state 
\cite{lee}.  With this in mind, we construct a simple model to describe our measurements of 
$n_{S}(T)$ and literature results for $\gamma(T)$, in which only portions of the FS unoccupied by 
the pseudogap contribute to the superfluid and entropy in the superconducting state.  

We present new measurements of $1/\lambda^{2}(T) \propto n_{S}(T)$ in 
YBa$_{2}$Cu$_{3}$O$_{7-\delta}$ (YBCO) films at various $\delta$.  We show that $1/\lambda^{2} (T)$ 
in optimally-doped films is that of a clean, strong-coupling d-wave superconductor with a full FS. 
Underdoped films are well described by a Fermi liquid-like model, in which electronic properties 
are expressed as integrals over the FS, but the integrals extend only over sections of the FS not 
occupied by the pseudogap in the normal state.  The fraction of the FS that survives is equal to 
the ratio of $n_{S}(0)$ of the underdoped film to $n_{S,opt.}(0)$ of the same film at optimal 
doping. 

The absence of critical fluctuations in optimally-doped YBCO films \cite{kmptc} and measurements of 
the effect of thermal phase fluctuations on 2D films of a conventional superconductor 
\cite{sjtmoge} lead us to conclude that fluctuation effects are weak in the underdoped films.  
However, the relative importance of thermal phase fluctuations (TPF's) to single particle 
excitations is controversial.  Carlson et al. \cite{eknew} have shown numerically that a 
fluctuation driven superfluid density in Josephson junction (JJ) arrays displays features similar 
to some very clean YBCO crystals, namely, $T$-linear behavior at low-$T$ \cite{hardy1}, $T_{C}$ 
roughly proportional to $n_S(0)$, and a wide critical region \cite{kamal3d,pasler}.  Terahertz 
measurements of the sheet conductance of BSCCO films also suggest a wide fluctuation region 
\cite{corson}.  On the other hand, theoretical analyses of TPF's in underdoped cuprates conclude 
that fluctuations are too weak at low-$T$ to account for the $T$-linear behavior of $\lambda (T)$ 
\cite{franzkwon,trlqm}.  Consistent with these results, our estimates indicate that the effects of 
TPF's are minor. 

It is not known why fluctuation effects near $T_{C}$ are weak in optimally-doped YBCO films 
\cite{kmptc} and some crystals \cite{sripure}, while appearing strong in other crystals 
\cite{kamal3d,pasler}.  Films are of high quality, based on their $T$-linear $\lambda(T)$ at 
low-$T$ and transition widths less than 1 K.  Evidently, critical fluctuations are sensitive to 
structural differences which affect neither of these quality indicators.  An estimate of the 
coupling between CuO bilayers in optimally-doped YBCO \cite{coupling} finds that for $T$ within 5 K 
of $T_{C}$, the ratio of interlayer coupling to in-plane coupling (J'/J in ref. \cite {eknew}) is 
greater than unity and fluctuations should be strongly suppressed.  To us, the presence of critical 
fluctuations in crystals is surprising, but their absence in films is not.  We speculate that a 
significant portion of what appear to be critical fluctuations in crystals is, in fact, due to the 
rapid decrease in the quasiparticle scattering rate as $T$ decreases below $T_{C}$ \cite{bonn} 
which serves to rapidly increase the conductivity, $\sigma_{1}(\omega,T)$, for $\omega$ less than 
the gap frequency, $\Delta_{0}(T)/\hbar$, thereby increasing $n_{S}$ \cite{tinkham}.  The decrease 
in scattering rate is known to be less rapid in disordered samples \cite{disorder}. 


The d-wave theory which we use is an extension of the weak-coupling result for $\lambda^{2}(0) / 
\lambda^{2}(T/T_{C0})$ \cite{wonmaki} to strong coupling by increasing the ratio 
$\Delta_{0}(0)/k_{B}T_{C0}$ above its weak-coupling value of 2.14, while preserving the dependence 
of $\Delta_{0}(T/T_{C0})/\Delta_{0}(0)$ on $T/T_{C0}$.  $T_{C0}$ is the mean-field transition 
temperature.  $1/\lambda^{2}(T)$ and $\gamma (T)$ are determined from the usual FS integrals over 
{\bf k}-space and energy. 

\fussy For simplicity, we neglect possible deviations of $\Delta ({\bf k},T)$ from 
$\Delta_{0}(T)$cos(2$\phi$) which are reported in ARPES measurements on underdoped 
Bi$_{2}$Sr$_{2}$Ca$_{1}$Cu$_{2}$O$_{8+\delta}$ (BSCCO) \cite{arpes} and which may or may not be 
present in YBCO.  Possible anomalous behavior of the quasi-1D CuO chains \cite{carbotte1} is not 
included.  Because such behavior is not observed in untwinned crystals, it is likely that chain 
specific effects are negligible in our highly twinned films.  Variation of the Fermi velocity, 
$v_{F}({\bf k})$, with $\bf k$ is not included.  If $n_{S}(0)$ is a factor $F$ smaller than in the 
optimally-doped film, then FS integrals are taken only over the angular interval, $\pm F(\pi /4)$, 
centered at each node in $\Delta({\bf k},T)$. 

2D TPF's are predicted to suppress $n_{S}(T)$ \cite{theoryfluc} by a factor: $a f_{Q}(T) k_{B} T 
\mu_{0} \lambda_{\perp}(T)/ \hbar R_{Q}$.  $1/\lambda_{\perp}(T) \equiv d/\lambda^{2}(T)$; $d$ = 
11.7 $\AA$ for YBCO.  The prefactor, {\it a}, is not well known theoretically.  We use $a = 0.175$ 
which is consistent with measurements on {\it a}-MoGe films \cite{sjtmoge} and calculations for a 
hexagonal array of resistively shunted JJ's \cite{trlqm,stroud}.  $R_{Q} \equiv \hbar/4e^{2} \sim 
1027 \; \Omega$ and $f_{Q}$ ($0 \leq f_{Q} \leq 1$) represents quantum suppression of TPF's.  The 
2D transition, $T_{2D}$, is expected where $\lambda_{\perp}(T_{2D}) T_{2D} \simeq \pi \hbar 
R_{Q}/2k_{B} $ = 9.8 mmK.  Since $T_{2D}$ is near $T_{C}$, this condition is approximately 
$\lambda_{\perp}(T_{2D}) \simeq 9.8 \; {\rm mm K} /T_{C}$.  (In the absence of quantum effects, 
$T_{2D}$ is the Kosterlitz-Thouless-Berezinskii transition temperature $T_{KTB}$.)  As $T$ 
approaches $T_{2D}$, the suppression of $n_{S}(T)$ grows rapidly due to nonlinear effects, reaching 
20\% to 50\% just below $T_{2D}$. 

Quantum suppression of TPF's in films has been controversial for some time.  Recently, quantum 
effects were predicted \cite{trlqm} and observed \cite{sjtmoge} to suppress TPF's when $k_{B}T/ 
\hbar$ drops below the "$R/L$" frequency of the film.  Approximately, the sheet inductance is $L = 
\mu_{0} \lambda_{\perp}(T)$ and the sheet conductance is $1/R = \sigma_{1}(\omega \lesssim 
\Delta_{0}/\hbar,T)d$.  For optimally-doped YBCO, $f_{Q}$ should be much less than unity below $0.8 
\; T_{C}$. 


Data presented here for films grown by coevaporation and sputtering are representative of other 
high quality films.  Films allow the rapid and reversible adjustment of oxygen content without 
altering the thickness or microstructure of the film.  The coevaporated YBCO films were grown using 
the BaF$_2$ method \cite{coevap} with a room temperature SrTiO$_{3}$ substrate in an atmosphere of 
about $5\times 10^{-6}$ torr of O$_{2}$, with a postanneal in wet oxygen.  After careful refinement 
of the growth rates from measurements of film stoichiometry by Rutherford Backscattering, YBCO 
films grown by this method consistently display a linear low-{\it T} penetration depth.  Additional 
films made by RF sputtering and by {\it in-situ} coevaporation of Sm, Ba, and Cu \cite{smba} onto 
substrates held at 750 C to 800 C are presented.  Optimally-doped film transition widths were less 
than 1 K, based on the peak in the real conductivity $\sigma_{1}(T)$.  One of the codeposited YBCO 
films was deoxygenated three times to $\sim$ O$_{6.8}$, O$_{6.7}$, and O$_{6.6}$ (determined from 
$T_{C}$) using low temperature anneals in argon. 
 
The complex conductivity, $\sigma = \sigma_{1} - i \sigma_{2}$, is determined from the mutual 
inductance of coaxial coils driven at 50 kHz located on opposite sides of the film \cite{sjtpd}. 
With a well defined coil geometry and known film thickness, $s$, Maxwell's equations provide 
$\sigma$ as a function of the 

\begin{figure}[htb] 
\centerline{\epsfxsize= 3.0in \epsfbox{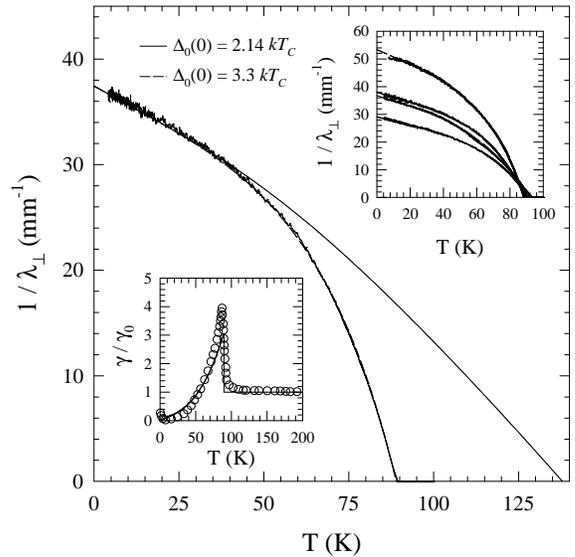}} \vspace{0.1in} \caption{ The inverse of the 2D 
magnetic penetration depth ($1/\lambda_{\perp} = d / \lambda^{2} \propto n_{s}$) for a YBCO film is 
shown with best low-$T$ fits using a weak coupling gap $\Delta_{0}(0)=2.14 \; k_{B}T_{C}$ (solid 
line) and strong coupling gap $\Delta_{0}(0)=3.32 \; k_{B}T_{C}$ (dashed line).  The data and the 
strong coupling fit are nearly indistinguishable.  The upper right inset shows four additional 
films with strong coupling mean-field fits $\Delta_{0}(0)=3.2 \pm 0.2 k_{B}T_{C}$ (dashed lines).  
The lower left inset compares the normalized electronic specific heat coefficient from Loram {\it 
et al.} [27] (open circles) with the d-wave result for $\Delta_{0}(0)=3.32 \; k_{B}T_{C}$ (solid 
line).} 
\end{figure}

\noindent measured mutual inductance.  Great care is taken to ensure that measurements are made in 
the linear response regime to within 0.1 K of $T_{C}$ by taking successive measurements at 
increasingly smaller drive coil currents.  $\sigma_{1}$ is very much smaller than $\sigma_{2}$ 
everywhere except very close to $T_{C}$.  From $\sigma_{2}$ we define $\lambda_{\perp}$: 

\ba \lambda_{\perp}(T) \equiv \frac{s/d}{\mu_{0} \sigma_{2} \omega}. \label{Lperp} \ea 

\noindent While $\lambda_{\perp}(0)$ is determined to an accuracy of about 10\%, limited by 
uncertainty in film thickness, relative changes induced by deoxygenation are known to better than 
5\%. 


Figure 1 shows $1/\lambda_{\perp} (T)$ for an optimally-doped YBCO film.  The solid curve is 
weak-coupling d-wave theory ($\Delta_{0}(0)/k_{B}T_{C0} = 2.14$) fitted to the low-$T$ data by 
adjusting $T_{C0}$.  The dashed curve, which is nearly indistinguishable from the data, is 
strong-coupling d-wave theory with $\Delta_{0}(0)/k_{B} = 3.3 \; T_{C0} \simeq 300$ K, consistent 
with tunneling measurements \cite{tunneling}.  The upper right inset to Figure 1 shows the 
excellent agreement between the measured $1/\lambda_{\perp}(T)$ and mean-field curves for several 
optimally-doped YBCO films and one film of \linebreak SmBa$_{2}$Cu$_{3}$O$_{7-\delta}$.  Given 
$1/\lambda_{\perp}(0)$ and taking $v_{F}$ = $0.7 \times 10^{5}$ m/s \cite{vf}, $\gamma(T)$ 
calculated with the same gap compares well with data \cite{loram1} on bulk YBCO (lower left inset 
to Fig. 1, with $\gamma_{0}$ = 16 mJ/mole  K$^{2}$).  The small experi-

\begin{figure}[htb]
\centerline{\epsfxsize= 3.0in \epsfbox{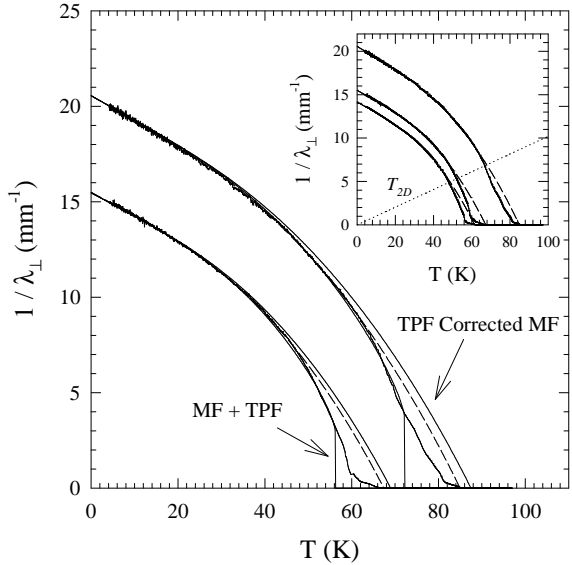}} \vspace{0.1in} \caption{The inset shows data 
for a coevaporated YBCO film at 3 stages of oxygen depletion, each successive stage having a lower 
$1 / \lambda_{\perp}(0)$.  Best low-$T$ extrapolations are given by $\Delta_{0}(0)\simeq 2.5 \; 
k_{B}T_{C}$ and are shown as dashed lines.  The intersection between the data and the dotted line 
is where the classical 2D phase fluctuation transition would occur for completely decoupled layers.  
The main figure displays the data from the first two deoxygenation steps and fits incorporating 2D 
phase fluctuations (MF+TPF) which show a discontinuous drop at $T_{2D}$.  The upper solid line are 
the corresponding curves with fluctuation effects removed (TPF Corrected MF).} 
\end{figure}

\noindent mental $\gamma (T)$ at low-$T$ could be fit better if the model allowed a {\bf 
k}-dependent $v_{F}$ and deviations of $\Delta({\bf k})$ from cos(2$\phi$). 

The inset to Figure 2 shows $1/\lambda_{\perp}(T)$ for a YBCO film at three stages of 
deoxygenation.  It is striking that $\partial n_{S}(T) / \partial T$ at low-$T$ (not necessarily 
the normalized \linebreak shape) is essentially independent of doping despite the introduction of 
oxygen vacancies into the CuO chains.  In our model, the unchanged slope, $d 
\lambda_{\perp}^{-1}(T)/dT |_{T \rightarrow 0}$, indicates no change in the opening of the gap near 
the gap nodes \cite{unchanged}.  The downward curvature of $1/\lambda_{\perp}(T)$ reflects that of 
$\Delta_0 (T)$ if TPF's are negligible.  We wish to extract the underlying mean-field behavior by 
estimating the effect of 2D TPF's.  For reference, the intersection of the dotted line with 
$1/\lambda_{\perp}(T)$ (inset to Figure 2) would locate $T_{2D}$ were there no interlayer coupling 
and no quantum suppression of phase fluctuations. 

The dashed curves in the inset and main portion of Figure 2 are extrapolations of the data taken 
below $T_{2D}$.  The lower solid curves are fits in which hypothetical mean-field behavior (upper 
solid curves) is suppressed by TPF's.  Since the influence of TPF's is determined by $T$, the 
normal-state sheet resistance, and $\lambda_{\perp}(T)$, all measured quantities, the only fitting 
parameter is $T_{C0}$.  The third deoxygenation step yields fits similar to the first two, but is 
left out for clarity.  A reduction in fluctuations as a result of interlayer coupling would bring 
the fluctuation-

\begin{figure}[htb]
\centerline{\epsfxsize= 3.32in \epsfbox{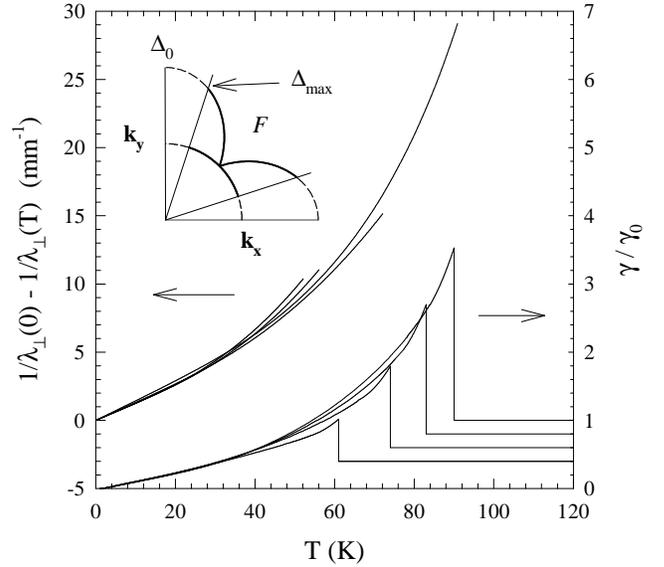}} \vspace{0.1in} \caption{ $n_{N}(T) \propto 
1/\lambda_{\perp}(0) - 1/\lambda_{\perp}(T)$ data for a YBCO film at optimal and 3 underdoped 
stages are shown.  The data are truncated above $T_{2D}$.  To the right are the calculated 
normalized electronic specific heat coefficients for $F$=1, 0.8, 0.6, and 0.4 (from highest peak to 
lowest peak).  {\bf k}-space integrals in the model, are taken only over a fraction of the Fermi 
surface, $F$, as illustrated in the inset. } 
\end{figure}

\noindent corrected mean-field result closer to the data, so uncertainty in mean-field behavior is 
bracketed by the upper solid curves and the data.  We take the simple extrapolations (dashed 
curves) as reasonable approximations to mean-field behavior at least for $T < T_{2D}$.  Our 
conclusions are insensitive to the extrapolation and to the "foot" seen just below $T_{C}$ which is 
likely due to oxygen inhomogeneity. 
 
Figure 3 displays the normal fluid density, $n_{N}(T) \propto 1/\lambda_{\perp}(0) - 
1/\lambda_{\perp}(T)$, represented by the mean-field (dashed) curves of Fig. 2.  Our model 
reproduces these when $\Delta_{0} (0) \simeq 300 K$ is fixed and $T_{C0} \simeq F^{2/5}$, where $F 
\equiv$ \linebreak $n_{S}(0)/n_{S,opt.}(0)$ in the experimental range, $0.4 \leq F \leq 1$.  The 
gap ratio, $\Delta_{0} (0)/k_{B}T_{C0}$, increases with underdoping, but the gap ratio defined from 
the maximum gap on the contributing FS segments, $\Delta_{max} = 
\Delta_{0}(0)cos(\frac{\pi}{2}(1-F))$, increases only slightly.  The contributing FS segments and 
$\Delta_{max}$ are shown pictorially in the inset to Figure 3. 

The model provides a basis for interpretation of $\gamma (T)$ \cite{loram1}.  For optimal and 
mildly underdoped YBCO, there are three striking features in the data.  One is that $\gamma(T)$ for 
$T \lesssim$ 60 K is nearly independent of doping.  Another is that the peak value of $\gamma(T)$ 
at $T_{C}$ decreases much more rapidly than $T_{C}$ with underdoping.  Finally, the electronic 
entropy just above $T_{C}$, {\it i.e.}, the integral of $\gamma(T)$ from 0 K to $T_{C}$, decreases 
significantly.  This implies that the hypothetical normal-state $\gamma(T)$ must decrease 
dramatically as $T$ decreases below $T_{C}$, even though $\gamma$ is nearly constant for $T > 
T_{C}$. 

Figure 3 shows that $\gamma(T)$ calculated with parameters fixed by $\lambda(T)$ has all of the 
experimental features mentioned above.  This is significant.  Simple models which account for the 
reduction in $n_{S}(0)$ by an increase in the effective mass of electrons, for example, would not 
describe $\gamma$ accurately.  Our model incorrectly predicts that $\gamma(T)$ just above $T_{C}$ 
should decrease with underdoping.  We hypothesize that our model describes the electron degrees of 
freedom that condense into the superconducting state, and that there exists, in addition, an 
anomalous contribution to $\gamma(T)$ which arises from degrees of freedom not associated with the 
superfluid (perhaps from electron spin degrees of freedom) and which decreases rapidly in the 
vicinity of $T_{C}$. 

For strong underdoping, $F \lesssim 0.4$, the model predicts that the thermodynamic critical field, 
$B_{C}(0)$, is proportional to $n_{S}(0)^{1/2} T_{C0}$, and the upper critical field, $B_{C2}(0)$, 
perpendicular to the $ab$-plane is proportional to $T_{C0}^{2}$ \cite{boyce}.


The authors would like to thank Aaron A. Pesetski and John A. Skinta for useful discussions and 
James E. Baumgardner II for numerical calculations and data analysis software.  The SmBaCuO film 
was generously provided by Vladimir Matijasevic. This work was supported in part by  DOE Contract 
No. DE-FG02-90ER45427 through the Midwest Superconductivity Consortium.


%
%
%
%


\begin{references}
\bibitem{coupling}  The indicator of $c$-axis coupling is $2\xi_{c}(T)/d$, where $\xi_{c}(T)$ is 
the $c$-axis coherence length.  Assuming $\xi_{c} \sim \xi_{c}(0)/(1 - T/T_{C0})^{\frac{1}{2}}$ 
near $T_{C}$ and $2\xi_{c}(0)/d = 2(\lambda_{ab}(0)/\lambda_{c}(0))(\xi_{ab}(0)/d) \simeq 1/3$ for 
optimally-doped YBCO, then at 0.9 $T_{C}$, $2\xi_{c}(T)/d \gtrsim 1$.
\bibitem{tinkham} The integral relationship between $\sigma_{1}$ and $n_{S}$ is described in: 
"Introduction to Superconductivity" by M. Tinkham, (Krieger, NY, 96).
\bibitem{vf}  A value twice as large as this is often quoted, but for the present purposes the 
difference is not important.
\end{references}
\end{document}